# A Formal Approach to System Integration Testing

Susanne Kandl
Institute of Computer Engineering
Vienna University of Technology
Austria
Email: susanne@vmars.tuwien.ac.at

Martin Elshuber
Institute of Computer Engineering
Vienna University of Technology
Austria
Email: martine@vmars.tuwien.ac.at

*Abstract*—System integration testing is the process of testing a system by the stepwise integration of sub-components. Usually these sub-components are already verified to guarantee their correct functional behavior. By integration of these verified sub-components into the overall system, *emergent* behavior may occur, i.e. behavior that evolves by the assembling of the sub-components. For system integration testing, both, the correct functional behavior of the overall system, and, the proper functioning of the sub-components in their system environment, have to be verified. In this work we present the idea of an approach for system integration testing based on formal verification. The system components are modeled in SystemC. In a first step these components are formally verified. Then a model of the overall system is built. In a second step this system model is formally verified. The novelty of this approach is given by two aspects: First, up to now the available verification frameworks for SystemC-models are more a proof of concept than really applicable to real industrial case studies. Secondly, although formal verification techniques are a common technique for the verification of software and hardware, by now they have only marginally considered for system integration testing.

## I. SYSTEM INTEGRATION TESTING

Testing (as part of the verification and validation of a system) is the process of checking whether the system under test behaves as defined by the specification. System integration testing is the process of testing the overall system by integrating sub-components. These sub-components have already been tested and verified before as self-contained systems. By integrating or assembling sub-components the overall system may show emergent behavior that evolves from the combination of the sub-systems. Consider following example: Component A represents a hardware architecture (ECU...electronic control unit) and was sufficiently verified in the hardware testing. The specification for the chip describes its functional behavior and the way how to use this hardware component. Component B is a software component. This component was developed and tested in a hardware-independent environment (e.g. SIL...software in the loop, i.e. running as a simulation on a PC). All the defined requirements of the software are verified. For integration testing the software is executed on the target-hardware architecture. Emergent effects are, for instance, that the proper functioning of the software depends on the memory management of the hardware (the chip). Conflicts in the management of hardware resources (e.g. memory) may cause failures in the software, although the software by itself is correctly implemented.

This work has been partially funded by the ARTEMIS Joint Undertaking and the National Funding Agency of Austria for the project VeTeSS under the funding ID ARTEMIS-2011-1-295311.

SystemC [1] is a de-facto industry standard for modeling systems at system level, and can be used to model software and hardware aspects in a single language. SystemC is an add-on library to C++ and provides constructs similar to Hardware Description Language (HDL) languages and a scheduler. Such models can be compiled to native machine code for most of the existing hardware architectures, thus allowing fast and accurate simulation of the system.

Although simulation is a proper method for detecting many bugs in a system, it cannot be used to verify whether a property of a system holds for every possible system state or not. Formal verification techniques can be used to guarantee the validity of a property for all possible system states.

## II. FORMAL APPROACH

### A. Ongoing work

Formal verification is a technique for verifying a system *property* for a system *model*. The system is represented as an automaton model, the property is formalized (e.g. as an LTL-formula...linear temporal logic) and a model verifier (e.g. a model checker) is used to verify whether the property is valid in the system model or not. Mainly two reasons make it difficult to formally verify SystemC models: (1) the constructs introduced by SystemC use a scheduler to execute and schedule processes activated on specific events. (2) SystemC allows to freely use C++ constructs like class inheritance, library functions, or Standard Template Library (STL). State-of-the-art techniques address these problems by transforming the model to another language and using existing tools. This transformation requires (ad 1) to model the scheduler explicitly, thus increasing the overall state space, and (ad 2) to restrict the SystemC model to a specific subset of SystemC (e.g. prohibiting class inheritance).

At the moment we are working on a framework for the formal verification of SystemC-models by transforming the SystemC-model into the formal language Promela, the input language for the model checker SPIN [2], see Figure 1. The gray blocks denote the steps that can be done by existing tools. These are SPIN, PinaVM [3] or LLVM [4]. Implementations for the white boxes are currently missing and are subject for ongoing work. Most steps (except Create Verifier C-Code and Compile & Link Verifier) are done by PinaVM or an extension of it. PinaVM itself is based on LLVM BC and uses LLVM libraries to handle the code. The output of PinaVM to the back-end is the LLVM BC, enriched with information on SystemC constructs, as well as the system architecture instantiated during model initialization. A prototype implementation of this tool is available [5].

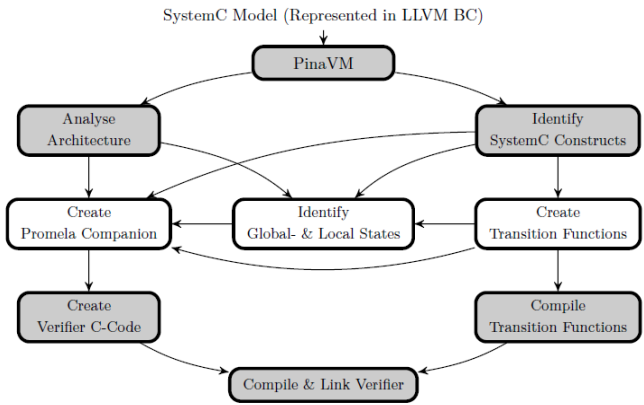

Fig. 1. Transformation from SystemC into Promela

### B. Further extensions

In future work we plan to extend this prototype for SystemC-verification by following two further features:

1. **A module-based verification feature:** In a first step the sub-components are modeled and verified. In the following a SystemC-model of the overall system is built. This model represents mainly the interactions between the sub-components. In a second step this system model can be verified by the verification framework with SPIN, see Figure 2. As the sub-components are already verified in a previous step, we can rely on the proper functioning of the sub-modules. For the step of integration we only have to consider the properties that emerge from assembling the sub-components. The idea is to concentrate on the modeling of the interfaces of the sub-components (e.g. relevant aspects of the hardware) and to derive the model for the integration testing process (at least partially) automatically. The main challenge will be to ensure the consistency between the models of the sub-components and the model for the overall system. As far as feasible we intend to *automate* parts of the derivation of the model for the overall system based on the given models of the sub-components. For this we have to find a procedure to *include* the aspects of a sub-component that are also relevant for the overall system and to *neglect* the details of the sub-component that have no impact on the overall system behavior. An analysis for an industrial use case showed that some components (components with clear defined interfaces) are easier to be considered for the system model than other components (e.g. firmware that is connected to all the parts of the hardware architecture). With this module-based verification approach for integration testing it is possible to detect possible integration problems already in the phase of the model-based development (and not after the final realization of the complete system).

2. **A test-case generator:** Test-case generation at the system level can be an intricate task. The test interfaces have to be re-defined and new test data has to be generated for the overall system. In general, only a small subset of the test data for the sub-components can be re-used for testing the overall system. We target to generate the test cases for the system automatically from the SystemC-model. The challenge for the realization of this technique is the mapping of the test cases derived from the SystemC-model onto the real system. This step is called concretization of test cases: The abstract test cases from the SystemC-model have to be adapted to the structure that can be executed in the real testing environment.

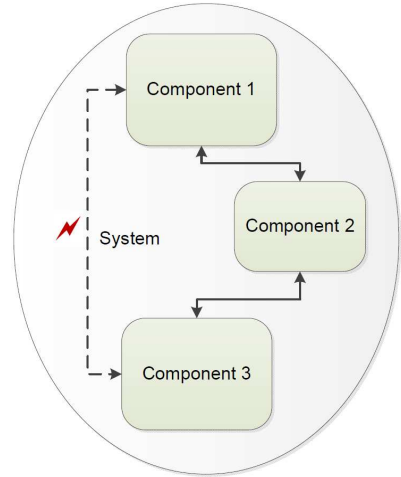

Fig. 2. Extension of the Verification Framework for Integration Testing

### III. RELATED WORK

An approach for the formal verification of SystemC is described in [6] and the referring proprietary tool Kratos [7]. Some approaches are based on the translation of the SystemC-model into the programming language C (by Scoot [8]) and applying model checkers for C, like CBMC [9].